\documentclass[sigchi]{acmart}

\usepackage{booktabs} 
\usepackage{balance}       

\copyrightyear{2019} 
\acmYear{2019} 
\setcopyright{iw3c2w3}
\acmConference[CHI 2019]{CHI Conference on Human Factors in Computing Systems Proceedings}{May 4--9, 2019}{Glasgow, Scotland Uk}
\acmBooktitle{CHI Conference on Human Factors in Computing Systems Proceedings (CHI 2019), May 4--9, 2019, Glasgow, Scotland Uk}
\acmDOI{10.1145/3290605.3300685}
\acmISBN{978-1-4503-5970-2/19/05}

\begin{document}
\title{Designing for Reproducibility: A Qualitative Study of Challenges and Opportunities in High Energy Physics}

\author{Sebastian S. Feger}
\affiliation{%
  \institution{LMU Munich and CERN}
  \city{Munich, Germany and Geneva, Switzerland}
}
\email{sebastian.s.feger@cern.ch}

\author{S{\"u}nje Dallmeier-Tiessen}
\affiliation{%
  \institution{CERN}
  \city{Geneva}
  \country{Switzerland}
}
\email{sunje.dallmeier-tiessen@cern.ch}

\author{Albrecht Schmidt}
\affiliation{%
  \institution{LMU Munich}
  \city{Munich}
  \country{Germany}}
\email{albrecht.schmidt@ifi.lmu.de}

\author{Pawe{\l} W. Wo{\'z}niak}
\affiliation{%
  \institution{Utrecht University}
  \city{Utrecht}
  \country{the Netherlands}
}
\email{p.w.wozniak@uu.nl}

\renewcommand{\shortauthors}{S. Feger et al.}

\begin{abstract}
Reproducibility should be a cornerstone of scientific research and is a growing concern among the scientific community and the public. Understanding how to design services and tools that support documentation, preservation and sharing is required to maximize the positive impact of scientific research. We conducted a study of user attitudes towards systems that support data preservation in High Energy Physics, one of science's most data-intensive branches. We report on our interview study with 12 experimental physicists, studying requirements and opportunities in designing for research preservation and reproducibility. Our findings suggest that we need to design for motivation and benefits in order to stimulate contributions and to address the observed scalability challenge. Therefore, researchers' attitudes towards communication, uncertainty, collaboration and automation need to be reflected in design. Based on our findings, we present a systematic view of user needs and constraints that define the design space of systems supporting reproducible practices.
\end{abstract}

%
%



\maketitle

\section{Introduction}

Reproducibility and reusability are core scientific concepts, enabling knowledge transfer and independent research verification. Alarming reports concerning the failure to reproduce empirical studies in a variety of scientific fields ~\cite{Baker2016, Prinz2011, Bonnet2011} are leading to the development of services, tools and strategies that aim to support key reproducible research practices ~\cite{Worden2017}.

\textit{Preserving} and \textit{sharing} research are basic requirements in reproducible science ~\cite{Bechhofer2013, Wilkinson2016, FAIR}, requiring efforts to describe, clean and document resources ~\cite{Borgman:1297241}. But those efforts are often not matched by the perceived gain. In fact, studies claim that the scientific culture does not support or even impairs compliance with reproducible practices ~\cite{Begley2012,Collaboration2012}.

As research preservation tools are emerging, we set out to study design requirements for technology that supports reproducible research practices. We studied data sharing and preservation flows and attitudes towards preservation systems in High Energy Physics (HEP), one of the most data-intensive branches of science \cite{Mélissa:2276551}. The volume of data and the community's demonstrated early adoption of computer-supported technology --- most notably the invention of the World Wide Web \cite{Berners-Lee1992} --- make for a strong environment to study technologies and strategies that are expected to become increasingly relevant in data-driven science; also referred to as the fourth paradigm of science \cite{Bell2006}.

We conducted our interview study with experimental physicists at CERN, a key HEP laboratory. The study was closely connected to a research preservation prototype service, tailored to CERN's major experiments. Based on our findings, we map practices around data sharing and chart challenges and opportunities involved in designing for research preservation and reproducibility.
This paper presents: (1) a detailed description of data preservation flows in world's leading data-intensive science environment; (2) six themes that describe user attitudes towards data presentation systems and (3) implications for designing systems that support reproducible science.

This paper is organized as follows. First, we review requirements and challenges of reproducible research and past efforts in designing for research communities. Next, we describe our study's context, in particular HEP and the prototype research preservation service. We then provide details of our interview study. Afterwards, we report on the six themes we identified: \textsc{Motivation}, \textsc{Communication}, \textsc{Uncertainty}, \textsc{Collaboration}, \textsc{Automation} and \textsc{Scalability}. Finally, we present implications for designing technology that supports reproducible research practices.

\section{Related Work}
In this section, we provide: (1) an overview of definitions, requirements, discussed incentive structures for reproducible research and reflect on discussions concerning the role of replication in HCI; and (2) review previous work in designing for scientific communities and research practices.

\subsection{Reproducibility}
Definitions of reproducibility and related terms vary between different disciplines ~\cite{Feitelson2015}. Leek and Peng ~\cite{Leek2015} define reproducibility \textit{``as the ability to recompute data analytic results given an observed dataset and knowledge of the data analysis pipeline.''} Feitelson ~\cite{Feitelson2015} stresses that reproducibility is not limited to simply recreating exactly the same experiment, but defines it as a \textit{``reproduction of the gist of an experiment: implementing the same general idea, in a similar setting, with newly created appropriate experimental apparatus.''}

The latter definition of reproducibility fits well to data analysis in HEP, characterized by statistically combining earlier experiment data with later run data. This data enrichment allows researchers to prove scientific concepts based on statistical probability. Since analyses might be based on experiment data captured over a range of several years, the former definition of reproducibility applies: analyses are not simply re-executed, but enriched and adapted to new input.

In this paper, we use the terms reproducibility and reproducible science. While it is important for us to refer to semantic discussions \cite{Gomez2010,Drummond2009,Feitelson2015} regarding reproducibility and related terms, like replicability and repeatability, we aim generally at environments in which researchers are encouraged to describe, preserve and share their work, in order to make resources \textit{re-usable} in the future.

\subsubsection{Description and Preservation are Requirements}

In order to enable the reproducibility of an experiment, researchers have to follow a set of practices ~\cite{Bechhofer2013, Borgman:1297241}. Those include documentation of all relevant analysis artefacts. In their paper, B\'{a}n\'{a}ti et al. ~\cite{Bánáti2015} classified several dependencies --- that have a direct impact on the reproducibility of experiments --- into three categories: infrastructural dependency, data dependency and job execution dependency. According to their work, reproducibility of computational studies requires to fully document the computational environments, and to ensure that all experimental resources remain accessible.

Chard et al. ~\cite{Chard2015} highlight the importance of data publication systems in data-intensive science. The authors stress the need to describe requirements for data publishing and illustrate that sharing on simple and basic network-accessible storages --- like a Dropbox folder --- is insufficient. They demand published data to be identifiable, described, preserved and searchable, motivating the need for dedicated data publication systems.

\subsubsection{Incentivizing Reproducible Practices}
Missing rewards and incentive structures have been identified as core contributors to the reproducibility challenge. Studies highlight that conferences and journals may encourage or demand publishing relevant experiment data as part of the publication process ~\cite{Belhajjame2014,Stodden2014}. Other incentive structures are based on monetary benefits. Russell ~\cite{russell2013if} demands funding agencies to reward scientists based on the reproducibility of their research. Rosenblatt ~\cite{Rosenblatt336ed5} highlights the collaborative agreements between universities and the industry. Companies could provide financial benefits for reproducible data, thus improving the overall quality of the research collaboration. Understanding better the role of incentives in reproducible research practices will also be key in designing technology that supports reproducibility.

\subsubsection{Replication in HCI}

In HCI it is common to refer to \textit{replication} of research. Wilson et al. \cite{Wilson2013} stress that novelty-driven research and diversity in HCI require discussing the place of replication in HCI. They describe four notions: \textit{Direct replication} to validate findings; \textit{Conceptual replication} refers to validity based on alternative approaches; \textit{Replicate \& Extend} means to reproduce prior research before making further investigations; and finally \textit{Applied Case Studies} refers to application of research findings in real world contexts.

In their paper '\textit{Is replication important for HCI?}', Greiffenhagen and Reeves \cite{Greiffenhagen2013} also stress the need to understand \textit{aims and motivations} for replication in HCI. They argue to distinguish between \textit{"what may be \textit{replicable} and what is actually \textit{replic-ated}."} While \textit{replicable} means that research \textit{in principal} can be replicated, \textit{replic-ated} marks research that \textit{has been replicated}. This distinction relates to the role of HCI in science, similar to \textit{"psychology's own debates around its status as a science (that) are also consonant with these foundational concerns of 'being replicable'"}. The authors highlight that \textit{"to focus the discussion of replication in HCI, it would be very helpful if one could gather more examples from different disciplines, from biology to physics, to see whether and how replications are valued in these."} In fact, as part of our study we aim to better understand the role and value of reproducibility in HEP. However, our study focuses on perceptions and design requirements for technology that supports reproducible research and is not designed to contribute \textit{directly} to discussions on the role of replication in HCI.

\subsection{Design for Supporting Research Practices}

Research has shown that the design of scientific tools profits from taking a human-centered approach, instead of studying only technical requirements ~\cite{Molin:2016:UDA:2971485.2971561} and that even small changes to the interface of analysis systems leads to adapted behavior of scientists ~\cite{Jianu:2012:ESU:2207676.2208704}. Given that impact, it is clear that successful service design requires involving domain experts ~\cite{Thomer:2016:CSS:2851581.2892549} in the process. In fact, improving research infrastructures, e.g. for collaborative data generation and reuse, requires \textit{"a deeper understanding of the social and technological circumstances"} \cite{Oleksik:2012:BDS:2145204.2145376}, motivating our researcher-centered study approach. 

In the context of research replicability, Mackay et al. \cite{Mackay2007} presented \textit{Touchstone}, an experiment design platform for HCI research on interaction techniques. The authors highlight that it is difficult to compare new techniques to the variety of existing ones, because of the effort needed to replicate those. Thus, comparison is often done only for one standard technique. The described platform allows to specify experiments and supports researchers with the evaluation process. Experiment designs and log data can be exported and imported, enabling reuse, replication and extension of research.

As sharing of research enables accessibility and improves visibility, studies ~\cite{2011AGUFMIN53B1628S,10.7717/peerj.175} found a clear connection between citation benefits for publications and open sharing of their experiment data. Thus, concerning the design of a community data system, Garza et al. ~\cite{Garza:2015:FCD:2783446.2783605} found that emphasizing \textit{``the potential of data citations can affect researchers' data sharing preferences from private to more open.''} And also badges have proven to encourage research sharing. Kidwell et al. \cite{Kidwell2016} compared contributions to the \textit{Psychological Science} journal, that adopted open science badges, to other journals in the same domain that have not done so. Papers got a visible badge in case data or materials from the reported study were released, leading to a significant increase in data sharing. ACM introduced very similar and even more fine-grained open research badges that even promote rewarded publications in their digital library \cite{boisvert2016incentivizing, acmbadgesweb}.

\section{Research Context}
We conducted our study at the European Organization for Nuclear Research (CERN). The study profited from the amount of data recorded in CERN's experiments, the demonstrated early adoption of computer-supported technology and an existing, tailored research preservation service.

\subsection{HEP, CERN and the LHC Collaborations}
In recent years, CERN received attention for discoveries surrounding the Large Hadron Collider (LHC). The LHC is the world's largest and most powerful particle accelerator ~\cite{Evans2008}. At four locations, particle collisions are measured by detectors, each of which is represented by a so-called LHC \textit{collaboration}. The four main LHC collaborations are: ALICE, ATLAS, CMS and LHCb ~\cite{Gustafsson2006}. To be able to verify findings, LHC collaborations mostly perform their research independently from others. As Cho ~\cite{Cho1564} highlights, that is especially true for CMS and ATLAS that have similar research goals, thus creating competition. Even though all research data are recorded locally within the detectors, LHC collaborations are not simply local organizational structures at CERN, but rather a global network that includes hundreds of institutes worldwide\footnote{https://greybook.cern.ch/greybook/researchProgram/detail?id=LHC}. However, despite their global scale, CERN is their center point. Concerning the structure of LHC collaborations, Merali ~\cite{merali2010large} argues that there is no simple top-down decision making, but rather a distribution of responsibility towards the many highly specialized teams. Merali further refers to a spokesperson who notes that "in industry, if people don't agree with you and refuse to carry out their tasks, they can be fired, but the same is not true in the LHC collaborations." That is because "physicists are often employed by universities, not by us." These are important aspects to consider in this study, as we can not rely on a central facilitator to command compliance with reproducible practices.

Despite competition between LHC collaborations, openness in scholarly communication is characteristic in HEP. The preprint server culture enables scientists to share ideas and results freely and immediately \cite{Gentil-Beccot2010,doi:10.1177/0306312716659373}. In her ethnographic study, Velden \cite{Velden2013} illustrates the openness that characterizes scholarly communication in HEP. She illustrates, how --- despite competition --- groups working with shared, large-scale facilities, share information in a relatively open fashion.

A pillar of the open research practices is the field's ability to develop and adapt to supportive technologies. It is not coincidental that the roots of the World Wide Web (WWW) lead back to CERN, where it was conceived to share data between institutes around the world \cite{Berners-Lee1992,birth:1998446, bentley1995supporting}. And still today, HEP makes for a strong environment to study handling of unmatched data volumes, as HEP remains to be one of the most data-intensive branches of science \cite{Mélissa:2276551}.

\subsection{CERN Analysis Preservation (CAP)}

\begin{figure}
  \includegraphics[width=1.0\columnwidth]{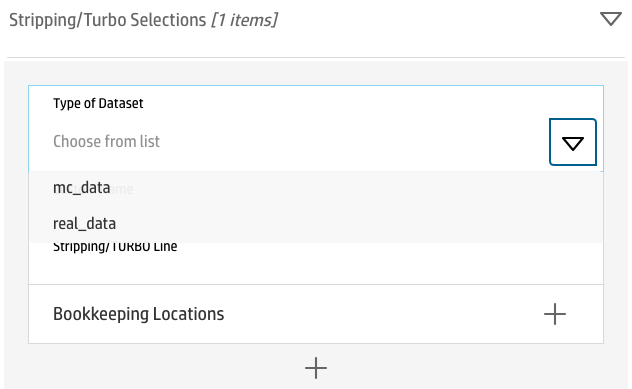}
  \Description{Shows a screenshot depicting part of an analysis documentation template on CAP. The screenshot illustrates how analysts are supported by the service through automation and suggestion. Here, a dropdown list limits applicable choices. The selection helps to further verify following inputs.}
  \caption{Part of the analysis submission form that allows physicists to describe and preserve their analyses. Supportive mechanisms ease efforts, ensure that data map to the internal LHC collaboration structures and guarantee consistency between records. In this scenario, researchers can chose between two possible types of datasets. Based on this choice, input in the following fields can be validated.}
  \label{fig:figure_cap}
\end{figure}

The CERN Analysis Preservation (CAP) prototype service\footnote{Publicly available on GitHub: 

\url{https://github.com/cernanalysispreservation}} enables researchers from the LHC collaborations to describe their analyses, consisting of data, metadata, workflows and code files ~\cite{chen2016cern}. Stored descriptions, data and files are preserved. The service thereby supports key reproducibility requirements: rich data description and long-term preservation. One of the key elements of CAP is a web-based graphical user interface that allows physicists to easily describe their analyses. Figure \ref{fig:figure_cap} shows a part of the LHCb analysis submission form. Due to differences in data analysis structures, analysis preservation templates are tailored to the experiment to which they belong. Initially, analyses on CAP are accessible to the creator as drafts. They can be shared with the whole LHC collaboration or individual collaboration members. Analyses are not shareable between different LHC collaborations.

The prototype is currently tested in a joint effort with several LHC collaborations. It is designed as a service that provides an easy and consistent way of describing and storing LHC analyses. Efforts were taken to support researchers in the description process. Depending on the data that are stored in the individual collaboration databases, CAP tries to auto-complete and auto-suggest as much information as possible. Nevertheless, the time required to fully describe and store an analysis is significant and adds to researchers' workload.

\section{Method}
We carried out 12 semi-structured interviews, to establish an empirical understanding of data sharing and preservation practices, as well as challenges and opportunities for systems that enable preservation and reproducibility.

\subsection{Recruitment and Participants}

In this section, we provide rich descriptions of the participants, including researchers' affiliations and experience levels. The analysts' ages ranged from 24 to 42 years old (average = 33, SD = 5.2). We decided not to provide information on the age of individual participants, as it would --- in combination with the additional characteristics --- allow to identify our participants. The 12 interviewees included 1 female (P8) and 11 males. The male oversampling reflects the employment structure at CERN: in 2017, between 79\% and 90\% (depending on the type of contract) of the research physicists working at CERN were male ~\cite{CERN-HR-STAFF-STAT-2017}. All interviewees were employed at CERN or at an institute collaborating with CERN. As all interviews were conducted during regular working hours, they became part of an analyst's regular work day. Accordingly, no additional remuneration was provided.

\begin{table}
  \centering
  \begin{tabular}{l c c r}
    {\small\textit{Interviewee reference}}
    & {\small \textit{Affiliation}}
    & {\small \textit{Gender}}
    & {\small \textit{Experience}}\\
    \midrule
    P1 & ATLAS & Male & Postdoc \\ 
    P2 & LHCb & Male & PhD student \\ 
    P3 & LHCb & Male & Senior researcher \\
    P4 & CMS & Male & Postdoc \\
    P5 & CMS & Male & Postdoc \\
    P6 & CMS & Male & Senior researcher \\
    P7 & CMS & Male & Senior researcher \\
    P8 & CMS & Female & PhD student \\
    P9 & CMS & Male & Convener \\
    P10 & CMS & Male & Senior researcher \\
    P11 & LHCb & Male & Convener \\
    P12 & CMS & Male & PhD student \\
  \end{tabular}
  \caption{Overview of the affiliations and professional experiences of the interviewees. We recruited data analysts from three LHC collaborations with a wide variety of experience. The male oversampling reflects the employment structure of research physicists at CERN.}~\label{tab:table_interviewees}
\end{table}

\subsubsection{Collaborations and Experience}

We interviewed data analysts working in three main LHC collaborations. Our recruitment focused on CMS and LHCb, as their preservation templates are most complex and developed. No interviewee had a hierarchical connection to any of the authors. Table \ref{tab:table_interviewees} provides an overview of the interviewees' affiliations with the LHC collaborations.

We selected physicists with a diverse level of experience and various roles to ensure a most complete representation of practices and perceptions. Half of the interviewees are early-stage researchers: PhD students and postdocs. The other half consists of senior researchers. As all interviewees - except the PhD students - held a PhD, we introduced metrics to distinguish between postdocs and senior researchers. In accordance with the maximum duration of postdoctoral fellowship contracts at CERN, we decided to consider as \textit{senior researchers} all interviewees who had worked for more than three years as postdoctoral physics researchers.

Two of the senior researchers had a convening role, or had such responsibilities within the last two years. Conveners are in charge of a working group and have a project management view. They are, however, often working on analyses themselves. Since they have this unique role within LHC collaborations, we identified them separately in Table \ref{tab:table_interviewees}.
 
\subsubsection{Cultural Diversity}

According to 2017 personnel statistics ~\cite{CERN-HR-STAFF-STAT-2017}, CERN had a total of 17,532 personnel, of which 3,440 were directly employed by the organization. CERN has 22 full member states, leading to a very diverse work environment. We decided not to list the nationalities of individual scientists, as several participants asked us not to do so and because we were concerned that participants could be identified based on the rich characterization already consisting of affiliation, experience and gender. However, we report the nationalities involved. The participants were in alphabetical order: British, Finnish, German, Indian, Iranian, Italian, Spanish and Swiss. The official working languages at CERN are English and French, with English being the predominant language in technical fields. All interviews were conducted in English. Working in a highly international environment at CERN, all interviewees had a full professional proficiency in English communication.

\subsection{Interview Protocol}

Initially, participants were invited to articulate questions and were asked to sign the consent form. The 12 interviews lasted on average 46 minutes (SD = 7.6). The semi-structured interviews followed the outline of the questionnaire:

Initially, questions targeted practices and experiences regarding analysis storage, sharing, access and reproducibility. Interviewees were encouraged to talk about expectations regarding a preservation service and the value of re-using analyses. This part of the questionnaire informed the themes \textsc{Motivation} and \textsc{Communication}. Next, we provided a short demonstration of the CAP prototype. Participants were introduced to the analysis description form and to collaborative aspects of the service: sharing an analysis with the LHC collaboration and accessing shared work. Participants were asked to imagine the service as an operational tool and were invited to describe the kind of information they would want to search for.

We used two paper exercises to support the effort of uncovering the underlying structure of analyses, as perceived by data analysts. In one exercise, participants were asked to design a faceted search for a search result page, showing a set of analyses with abstract titles. They had three empty boxes at their disposal and could enter a title and four to seven characteristics each. In the second exercise, we encouraged participants to draw connections and dependencies that can exist between analyses on a printout with two circles, named \textit{Analysis A} and \textit{Analysis B}. The exercise supported us in understanding the value of a service being aware of relations between analyses. Finally, interviewees were encouraged to reflect on CAP and invited to describe how they keep aware of colleagues' ongoing analyses within their LHC collaboration.

The system-related part of the questionnaire and the paper exercises informed our results about \textsc{Uncertainty}, \textsc{Collaboration} and \textsc{Automation}.

\subsection{Data Analysis}

All interviews were transcribed non-verbatim by the principal author. We used the Atlas.ti data analysis software to organize, code and analyze the transcriptions. Thematic analysis \cite{Blandford:2222613} was used to identify emerging themes from the interviews. We performed an initial analysis after the first six interviews were conducted. At first, we repeatedly read through the transcriptions and marked strong comments, problems and needs. Already at this stage, it became apparent that analysts were troubled by challenges the currently employed communication and analysis workflow practices posed. After we got a thorough understanding of the kind of information contained in the transcriptions, we conducted open coding of the first six interviews. As the principal author and two co-authors discussed those initial findings, we were content to see the potential our interviews revealed: the participants already described tangible examples of how a preservation service might motivate their contribution as a strategy to overcome previously mentioned challenges. We decided not to apply any changes to the questionnaire.

As the study evolved, we proceeded with our analysis approach and revised already existing codes. We aggregated them into a total of 34 code groups that were later revised and reduced to 22 groups. The reduction was mainly due to several groups describing different approaches of communication, learning and collaboration. For example, three smaller code groups that highlighted various aspects of e-mail communication were aggregated into one: \textit{E-Mail (still) plays key role in communication}. We continued to discuss our evolving analysis while conducting the remaining interviews. In addition, the transcript of the longest interview was independently coded by the principal author, one co-author and one external scientist, who gained expertise in thematic content analysis and was not directly involved in this study.

A late version of the paper draft was shared with the 12 interviewees and they were informed about their interviewee reference. We encouraged the participants to review the paper and to discuss any concerns with us. Eight interviewees responded (P2, P4, P5, P7, P8, P9, P11, P12), all of which explicitly approved of the paper. We did not receive critical comments regarding our work. P9 provided several suggestions, almost all of which we integrated. The CMS convener also proposed to \textit{"argue that the under-representation of ATLAS is not a big issue, as it is likely that the attitudes in the two multi-purpose experiments are similar (the two experiments have the same goals, similar designs, and a similar number of scientists)."}

\section{Findings}

Six themes emerged from our data analysis. In this section, we present each theme and our understanding of the constraints, opportunities and implications involved.

\subsection{Motivation}

Our analysis revealed that personal motivation is a major concern in research preservation practices. In particular P1, P2, P7, P9 and P11 worry about contribution behaviors towards a preservation service. P1 further contrasts information \textit{use} and \textit{contribution}: \textit{"People may want to use information - but we need to get them to contribute information as well."} The analyst calls this \textit{"the most difficult task"} to be accomplished. 

Several analysts (P1, P2, P9, P11) point to missing incentives as the core challenge. They stress that preserving data is not immediately rewarding for oneself, while requiring substantial time and effort. P9 highlights that even though analysts who preserve and share their work might get slightly more citations, this is \textit{"a mild incentive. It's more motivating to start a new analysis, other than spending time encoding things..."}

In this context, convener P11 critically contrasted policies with resulting preservation quality and highlighted the motivational strength of returned benefits: 

\smallskip

\noindent \textit{"...if you take this extra step of enforcing all these things at this level, it's never going to get done. Because if you use this as a documentation, so I'm done, now I'm going to put these things up. If it complains, like, I don't care... [...] But if there is a way of getting an extra benefit out of this, while doing your proper preservation, that is good - that would totally work."}

\smallskip

\noindent Imagining a service that not only provides access to preserved resources, but allows systematic execution of those, the convener states that he does not \textit{"see any attitude problem anymore, because doing this sort of preservation gives you an advantage."} Such immediate mechanisms might also provide incentives to integrate a preservation service into the analysis workflow, which according to P9 will be crucial. The convener expects that researchers \textit{"will not adapt to data preservation afterwards. Or five percent will do."}

\subsection{Communication}
Our analysis revealed that data analysts in HEP have a high demand for information. Yet, communication practices often depend on personal relations. All of our interviewees described the need to access code files from colleagues or highlighted how access could support them in their analysis work. Even though most analysts (P2 - P4, P6 - P8, P10 - P12) explicitly stated that they share their work on repositories that provide access to their LHC collaboration, information and resource flow commonly relied on traditional methods of communication: 

\smallskip

\noindent \textit{"The few times that I have used other people's code, I think that...I think it was sent to me by e-mail all the times"} (P3)

\smallskip

\noindent \textit{"They have saved their work and then I can ask them: 'where have you located this code? Can I use it?' And they might send me a link to their repository."} (P8)

\smallskip

\noindent The analysis of our interviews revealed the general practice of engaging in personal communication with colleagues in order to find resources. P4 highlights a common statement, i.e. colleagues pointing to existing resources:

\smallskip

\noindent \textit{"You go to the person you know is working on that part and you ask directly: 'Sorry, do you know where I can find the instructions to do that?' and he will probably point to the correct TWiki or the correct information"}

\smallskip

\noindent Personal relations are vital in this communication and information architecture. Most analysts (P1, P2, P3, P4, P6, P7, P8, P9, P11) stressed that it was important to know the right people to ask for information. P8 described the effort needed:

\smallskip

\noindent \textit{"I mean you have to know the right people. You have to know the person who maybe was involved in 2009 in some project. And then you have to know his friend, who was doing this. And his friend and then there is somebody who did this and she can tell you how it went."}

\smallskip

\noindent But, communication and information exchange was often contained within groups and institutes. P7 stressed that for a certain technique, other groups \textit{"have better ideas. In fact, I know that they have better ideas than other groups, but they are not using them, because we are not talking to each other."} P2 stated that \textit{"being shy and not necessarily knowing who to e-mail"} are personal reasons not to engage in communication with colleagues. The challenge to find the right colleagues to talk to is increased by the high rotation of researchers, many of them staying only few years.

Almost all analysts (P1 --- P4, P6 --- P11) in our study referred to another common issue they encounter: the lack of documentation. P6 illustrated the link between missing documentation and the need to ask for information instead:

\smallskip

\noindent \textit{"This is really mouth-to-mouth how to do this and how to do that. I mean the problem for preservation is that at the moment it's just: ask your colleague, rather than write a documentation and then say 'please read this.'"}

\smallskip

\noindent Meetings and presentations are a key medium in sharing knowledge. However, the practice of considering presentations as a form of knowledge documentation makes access to information difficult:

\smallskip

\noindent \textit{"There are cases you asked somebody: 'but did they do this, actually?' And somebody says like: 'I remember! Two years ago, there was this one summer meeting. We were having coffee and then they showed one slide that showed the thing.' And this slide might have never made it to the article." (P8)}

\subsection{Uncertainty}

\begin{figure}
\centering
  \includegraphics[width=1.0\columnwidth]{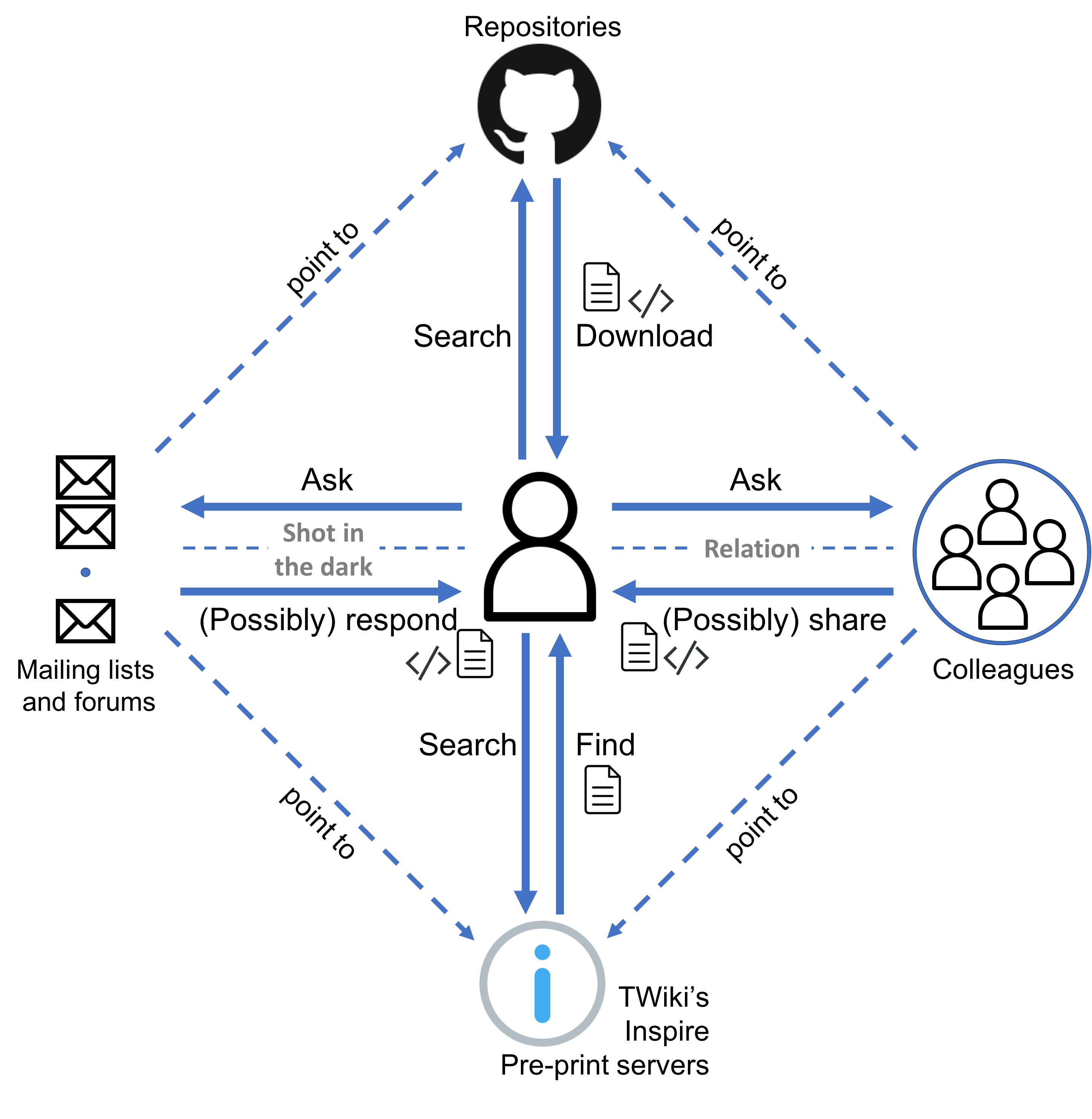}
  \Description{Illustrates the information and resource flow in HEP. At the center, a researcher needs information or resources. She / He can ask colleagues directly or on mailing lists. The outcome of these approaches is characterized by a high degree of uncertainty and relies on the strength of the personal network. Alternatively, the researcher can directly search for required information on repositories and / or on knowledge databases. However, navigating those can be cumbersome and finding information directly is difficult as well.}
  \caption{The current flow of information in HEP data analysis is characterized by the need to ask colleagues and the uncertainty of finding required resources.}~\label{fig:figure_current_flow}
\end{figure}

Our interview findings revealed that the communication and information architecture leads to two types of \textit{uncertainty}: (1) related to the \textit{accessibility} of information and resources; and (2) connected to the \textit{volatility} of data. 

\subsubsection{Accessibility}

\noindent As depicted in Figure \ref{fig:figure_current_flow}, analysts follow two principal approaches to \textit{access} information and resources: they search for them on repositories and databases or ask colleagues. The outcome of directly searching for resources contains uncertainty, as researchers might not be sure exactly what and where to search. But, also various search mechanisms represent challenges. A researcher described searching for an analysis and highlights, that \textit{"at the moment, it's sometimes hard to find even the ones that I do know exist, because I don't know whether or not they are listed maybe under the person I know. So, [name] I know that I can find... Well, actually I don't know if I can find his analysis under his GitHub user."} (P2)

Our interviewees (P1 --- P4, P6 --- P9, P11, P12) reported that they typically contact colleagues or disseminate requests on mailing lists and forums to ask for information and resources. While mailing lists represent a shot in the dark, the success of approaching colleagues is influenced by personal relations. If successful, they receive required resources directly or are pointed to the corresponding location.

\subsubsection{Volatility}

Facing vast amounts of data and dependencies, analysts wish that a centralized preservation service helps them with uncertainty that is caused by the \textit{volatility} of data.

\smallskip

\textit{Analysis Integrity:} A service aware of analysis dependencies can ensure that needed resources are not deleted.

\smallskip

\noindent \textit{"...and this can be useful even while doing the analysis, because what happens is that people need to make disk space and then they say: 'ah, we want to remove this and this and this dataset - if you need it, please complain.' And if you had this in a database for example, it could be used also saying like 'ah, this person is using this for this analysis' even before you would share your analysis." (P6)}

\smallskip

\noindent The analyst even highlighted the possibility to track datasets of work in progress that was not yet shared with the LHC collaboration. A convener also motivates the issue that comes with the removal of data and describes the effort and uncertainty involved in current communication practices:

\smallskip

\noindent \textit{"Sometimes versions get removed from disk [...] And the physics planning group asks the conveners: 'ok, is anybody still using those data?' [...] I have to send an email of which version they are using etc. [...] And at some point, if I have 30 or 40 analyses going on in my working group, it's very hard not to make a mistake in this sense if people don't answer the emails. While if I go here, I say ok, this is the data they are using - I know what they are using - and it takes me ten minutes and I can have a look and I know exactly." (P11)}

\smallskip

\textit{Receiving vital analysis information}: We learned that different analyses often have input datasets in common. When an analyst finds issues with a dataset, she or he draws back to the existing communication architecture.

\smallskip

\noindent \textit{"I present it in either one of the meetings which is to do with like that area of the detector for example. Or if it was something higher profile than maybe one of the three or four meetings which are more general, applicable to the collaboration\footnote{The interviewee is referring to the LHC collaboration.}. And from there that would involve talking to enough people in the management and various roles...that it would then I guess propagate to...they would be again in touch with whoever they knew about that might be affected." (P2)}

\smallskip

\noindent The risk of relying on this communication flow is that one might naturally miss vital information. An analyst could be unavailable to attend the right meeting or generally not be part of it. The person sending the email might also not know about all affected analyses. This might especially be true for relevant analyses that are conducted in a working group different from the ones of the analysts that are signaling the issue. A preservation service enabling researchers to signal warnings associated with a dataset or, generally, resources that are shared by various analyses, allows informing dependent analysts in a reliable manner. As being informed about discovered issues can be vital for researchers, it would be in their very interest to keep their ongoing analyses well documented in the service.

\smallskip

\textit{Staying Up-to-Date}: Keeping up-to-date on relevant changes can be challenging in data-intensive environments. Researchers hope that a preservation service provides reliable dependency awareness to analysts who document their work:

\smallskip

\noindent \textit{"The system probably tells me: 'This result is outdated. The input has changed'. Technical example. At the moment, this communication happens over email essentially" (P6)}

\smallskip

\noindent P11 told us about a concrete experience:

\smallskip

\noindent \textit{"He was using some number, but then at some point the new result came out and he had not realized. Nobody realized. And then, of course, when he went and presented things he was very advanced, they said 'well, there is a new result - have you used this? No, I have not used it.'"}

\subsection{Collaboration}

Sharing their work openly, analysts increase their chance to engage in collaboration. Currently, useful collaboration is hindered by missing awareness of what others do. We can imagine this to be especially true outside of groups and dislocated institutes. P4 emphasizes the value of collaboration:

\smallskip

\noindent \textit{"The nTuple production is a really time consuming part of the analysis. So, if we can produce one set of nTuples...so one group produces them and then they can be shared by many analysis teams...this has, of course, a lot of benefits."}

\smallskip

\noindent Researchers who document their ongoing activities and interests increase their discoverability within the LHC collaboration. Thereby, they increase their chance to be asked to join an official request that might satisfy their data needs:

\smallskip

\noindent \textit{"I want to request more simulation. [...] I would search and I would say these are the people. I would just write to them, because I want to do this few modifications. But maybe this simulation is also useful for them, so we can just get together and get something out." (P11)}

\smallskip

\noindent  In fact, a convener stated that due to the size of LHC collaborations, it is difficult to be aware of other ongoing analyses:

\smallskip

\noindent \textit{"CMS is so big that I cannot know if someone else is already working on it. So, if this tool is intended to have also the ongoing analyses since a very early stage, this would help me if I can know who is working on that." (P9)}

\smallskip

\noindent P8 highlights that being aware of other analyses can possibly lead to collaboration and prevent unwanted competition:

\smallskip

\noindent \textit{"Because the issue at CMS - and probably at whole CERN - is that you want start working on it, but, on the other hand, it's rude if you start working on something and you publish and then you get an angry message, saying: 'hey, we were just about to publish this, and you cannot do it.' [...] The rule is that everyone can study everything, but, of course, you don't want to steal anybody's subjects. So, if it wouldn't be published, you would then maybe collaborate with them."}

\subsection{Automation}

We see an opportunity to support researchers based on the common structure that applies to analyses: \textit{"because in the end, everybody does the same thing"} (P7). A convener characterized this theme by demanding \textit{"more and more Lego block kind analyses, keeping to a minimum the cases where you have to tailor the analysis a bit out of the path"} (P9).

\subsubsection{Templated analysis design}

As P11 articulates, the common steps and well-defined analysis structure represent an opportunity to provide checklists and templates that facilitate analysis work:

\smallskip

\noindent \textit{"If, of course, I have some sort of checklist or some sort of template to say 'what is your bookkeeping queries --- use this and that', then of course this would make my life easier. Because I would be sure I don't forget anything."}

\smallskip

\noindent The convener makes two claims on how a structured analysis description template could support researchers. First, templates help in the analysis design. Second, the service could inform about missing fragments or display warnings based on a set of defined checks. However, it is important to recognize a core challenge that comes with well-structured analysis templates; allowing for sufficient flexibility:

\smallskip

\noindent \textit{"Somehow these platforms tend to --- which is one of the strong points, but at the same time one of the weaknesses --- is that [...] it gives you some sort of template and makes it very easy for you to fill in the blanks. But at the same time, this makes things difficult, if you want to make very complex analyses where it's not so obvious anymore what you want to do." (P11)}

\smallskip

\subsubsection{Automate Running and Interpretation}

Several analysts (P2, P5, P7, P8, P11) expressed their wish for centralized platforms to automate tasks that they would currently have to perform manually. P2 stated:

\smallskip

\noindent \textit{"So, being able to kind of see that it...might be able to submit to it and then it just goes through and runs and does everything...and I don't need to think too much about whether or not something is going to break in the middle for something that is nothing related to me, would potentially be quite nice."}

\smallskip

\noindent However, not only automating the full execution of analyses seems desirable, but also interpretation of systematics:

\smallskip

\noindent \textit{"And I say: 'ok, now I want to know for example, which are the systematics' and you can tell me, because you know you have the information to do it by yourself. You will save a lot of time. People will be very happy I think." (P5)}

\smallskip

\subsubsection{Preventing mistakes}
P7 described how the similarity and common structure of analyses supports automated comparison and verification:

\smallskip

\noindent \textit{"What I would like to search is the names of the Monte Carlo samples used by other analyses. [...] the biggest mistake you can make is to forget one. Because if you forgot one, then you will see new physics, essentially. And it's a one-line mistake."}

\smallskip

\noindent Developing a feature that compares a list of dataset identifiers and that points to irregularities is trivial. Yet, as P7 continues to describe the effort needed to do the comparison at the moment, the perceived gain seems to be high:

\smallskip

\noindent \textit{"So, the analysis note always contains a table - it's a PDF. Then always contains a table with a list of Monte Carlos. I often download that, look at the table and see what's missing. Copy paste things from there. But so here, I would be able to do it directly here." }

\subsection{Scalability}

Although not directly in the scope of the questionnaire, four interviewees (P3, P8, P9, P11) commented on the growing complexity of analysis work in HEP, stressing the importance of preservation and reproducibility. P9 highlights the issues that evolve from collecting more and more data: 

\smallskip

\noindent \textit{"As we collect the data, the possibility of analysis grows. In fact, we are more and more understaffed, despite of being so many in the collaboration\footnote{The interviewee is referring to the LHC collaboration.}. Because, what is interesting for the particle physics community grows as data grow. And so, we get thinner and thinner in person power in all areas that we deem crucial."}

\smallskip

\noindent The convener adds that \textit{"a typical analysis cycle becomes much much longer. Typical contract duration stays the same."} P3 details how the high amount of rotation and (ir-)reproducibility impact analysis durations:

\smallskip

\noindent \textit{"If someone goes and an analysis is not finished, it might take years. Because there was something only this person could do. I think that analysis preservation could help a lot on this. [...] But otherwise you might have to study analyses from scratch if someone important disappears."}

\smallskip

\noindent P11 agrees that \textit{"it's getting more and more complex, so I think you really need to put things together in a way that is reasonable and re-runnable in some sort of way."} P9 coined the term \textit{orphan analyses}. It describes analyses for which no one is responsible anymore. The convener expects that \textit{"at some point it will become a crisis. Because, so far, it was a minority of cases of orphan analyses. It will become more and more frequent, unless contract durations will change. But this will not happen."}

\section{Implications for Design}
We present challenges and opportunities in designing for research preservation and reproducibility. Our work shows that the ability to access documented and shared analyses can profit both individual researchers and groups \cite{falessi2006documenting}. Our findings hint towards what Rule et al. \cite{rule2018exploration} call \textit{"tension between exploration and explanation in
constructing and sharing"} computational resources. Here, we primarily learned about the \textbf{need to motivate and incentivize contributions}. Based on our findings, we show how design can create motivating, \textit{secondary usage forms} of the platform and its content, related to uncertainty, collaboration and structure. And, while references in this section underline that the CHI community has established a long tradition of studying collaboration and communication around knowledge work, it is not yet known how to design collaborative systems that foster reproducible practices and incentivize preservation and data sharing. The following description of \textit{secondary usage forms} aims to contribute to knowledge about motivations and incentives for platforms that support research reproducibility.

\subsection{Exploit Platforms' Secondary Functions}

As observed in the \textsc{Motivation} theme, getting researchers to document and preserve their work is a main concern. In this context, researchers critically commented on the impact of policies, creating little motivation to ensure the preservation quality beyond fulfilling formal requirements. And also citation benefits, commonly discussed as means to encourage research sharing \cite{10.7717/peerj.175}, might provide only a mild incentive, as time required for documenting and preserving can be spend more rewarding on novel research. This seems especially true in view of growing opportunities that result from the increasing amount of data, as described in the \textsc{Scalability} theme. Yet, researchers indicated how centralized preservation technology can uniquely benefit their work, in turn creating motivation to contribute their research. Thus, we have to \textbf{study researchers' practices, needs and challenges in order to understand how scientists can benefit from centralized preservation technology. Doing so, we learn about the \textit{secondary function} of the platform and its content, crucial in developing powerful incentive structures}. 

\subsection{Support Coping with Uncertainty}
As we learned in the \textsc{Communication} theme, the information architecture is heavily relying on personal connections and communication, leading to a high degree of \textsc{Uncertainty} related to the \textit{accessibility} and \textit{volatility} of information and data. Consequently, researchers report encountering severe issues related to the insufficient transparency and structure that a centralized preservation service might be able to mitigate. We propose two strategies: First, a centralized preservation service can implement overviews and details of analysis dependencies not available anywhere else. Implementing corresponding features enables us to \textbf{promote preservation as effective strategy to cope with uncertainty} so that research integrity of documented dependencies can be guaranteed. Second, we further imagine documenting analyses on a dedicated, centralized service to be a powerful strategy to \textit{minimize} uncertainty towards updated dependencies and erroneous data, if the service provides awareness to researchers. In the case of data-related warnings, reliable notifications could be sent to analysts who depend on collaboration-wide resources, replacing current, less reliable communication architectures. This approach also relates to uncertainties at the \textit{data layer}, as described by Boukhelifa et al. \cite{Boukhelifa2017}, who studied types of uncertainty and coping strategies of data workers in various domains. According to their work, the three main active coping strategies are: \textit{Ignore}, \textit{Understand} and \textit{Minimize}. In summary, our findings suggest that such secondary benefits might drive researchers to contribute and use the preservation tool.

\subsection{Provide Collaboration-Stimulating Mechanisms}
The \textsc{Collaboration} theme highlighted the importance of cooperation in HEP. Analysts save time when they join forces with colleagues or groups with similar interests. Yet, awareness constraints resulting from the communication and information architecture often hinder further collaboration. We postulate that the preservation platform can add useful secondary benefits for theses cases. First, given the centralized interface and knowledge aggregation function of a preservation service, we see opportunities to \textbf{support locating expertise in research collaborations}. In fact, especially knowledge-intensive work profits from such supporting tools, as it enables sharing expertise across organizational and physical barriers \cite{Cross2004}. Ehrlich et al. \cite{ehrlich2007searching} note that awareness of "who knows what" is indeed key to stimulating collaboration. In an organizational context, Transactive Memory Systems (TMS) are employed to create such awareness. HEP collaborations are TMS, in that the sum of knowledge is distributed among their analysts and the communication between them forms a group memory system \cite{wegner1987transactive}. Further research on the support and integration of TMS in the context of platforms for research reproducibility could increase acceptance through heightened awareness provided by such platforms. Also, elements of social file sharing could further stimulate discovery and exploration of relevant researchers and analyses. As noted by Shami et al. \cite{shami2011browse}, this can be particularly important in large organizations.

Second, an important benefit could be the visibility of team or project members. Taking preserved research as basis for expertise location can incentivize contributions, as scientists who document in great detail are naturally most visible, thus increasing their chances to engage in collaboration. This approach also enables us to mitigate privacy concerns, by considering only resources of analyses that have been shared with the LHC collaboration. Mining documented and shared research to provide expertise location thus mitigates common challenges: Typically, workplace expertise locators infer knowledge either by mining existing organizational resources like work emails \cite{Campbell2003, Gopalakrishnan2017}, or by asking employees to indicate their skills and connections within an organization \cite{Shami2007}. While automated mining of resources may cause privacy concerns, relying on users to undergo the effort of maintaining an accurate profile is slower and less complete \cite{Reichling2009}. Given the increased interdisciplinary and international research culture, developing such bridging mechanisms --- even though not central to the service missions --- is especially helpful.

\subsection{Support Structured Designs}
A community-tailored research preservation service can support analysts through automated mechanisms that make use of prevalent workflow structures. Researchers pointed out that analysis work within a LHC collaboration commonly follows general patterns, demanding even to further streamline processes as much as possible; thereby pointing to the guiding role of preservation technology. We propose to \textbf{design community-tailored services that closely map research workflows to preservation templates}. That way, preservation services can provide checklists and guidance for the research and preservation process; furthermore, automation of common workflow steps can increase efficiency. Additionally, if the preservation service is well embedded into the research workflows, it could enable supportive mechanisms like auto-suggest and auto-completion. Such steps are key to minimizing the burden of research preservation, which is of great importance, as we acknowledge that the acceptance and willingness to comply with reproducible practices will always be related to the cost/benefit ratio related to research preservation and sharing.  
Having noted the need for automation and taylorization of interfaces, we need to emphasize the significance of academic freedom when designing such services. Design has to account for all the analyses, also those that are not reflected in mainstream workflows. We have to \textbf{support creativity and novelty by leaving contributors in control}. This applies both for supportive mechanisms like auto-complete and auto-suggest, as well as for the template design.

\section{Discussion}

The study's findings and implications have pointed to several relationships that are important for designing technology that enables research preservation and reproducibility. First, we have contrasted required efforts with returned benefits. It is apparent that stimuli are required to encourage researchers to conduct uninteresting and repetitive documentation and preservation tasks that in itself, and at least in the short run, are mostly unrewarding. Thus, not surprisingly, the call for policies is prominent in discussions on reproducible research. Yet, our findings hint towards the relation between preservation quality and policies, raising doubts that policies can encourage sustained commitment to documentation and preservation beyond a formal check of requirements. In this context, we hypothesize that also the relation between policies and flexibility needs to be considered. Thinking about structured description mechanisms as provided by CAP, one needs to decide on a common denominator that defines the main building blocks to comply with the policies. However, this is likely to create two problems: (1) Lack of motivation to preserve fragments that are not part of the basic building blocks of research conducted within the hierarchical structure for which the policies apply; (2) Preservation platforms that map policies might discourage or neglect research that is not part of the fundamental building blocks.  

Facing those conflicting relationships, meaningful incentive structures could positively influence the reproducibility challenge and create a favorable shift of balance between required efforts and returned benefits. We postulate that communities dealing with the design of such systems need to invest a significant amount of time into user research to create tailored and structured designs. Further research in this area is surely needed, i.e. the evaluation of prototypes or established systems in general and with a focus on the users' exploitation of secondary benefits of the system. This call for more research in this area is particularly evident when looking at the latest study by Rowhani-Farid et al. \cite{Rowhani-Farid2017} who  found only one evidence-based incentive for data sharing in their systematic literature review. They conducted their study in search of incentives in the health and medical research domain, one of the branches of science that was in the focus of reproducibility discussions from the very beginning. The only reported incentive they found relates to open science badges that resulted in a significant impact in data sharing of papers submitted to the \textit{Psychological Science} journal. The authors highlight that \textit{"given that data is the foundation of evidence-based health and medical research, it is paradoxical that there is only one evidence-based incentive to promote data sharing. More well-designed studies are needed in order to increase the currently low rates of data sharing."}

Our study showed how design can create secondary usage forms of preservation technology and its content related to communication, uncertainty, collaboration and automation. Described mechanisms and benefits apply not only to submissions at the end of the research lifecycle, but, rather, provide certainty and visibility for ongoing research. The significance of such contribution-stimulating mechanisms is particularly reflected in the observed scalability challenge, indicating that reproducibility in data-intensive computational science is not only a scientific ideal, but a hard requirement. This is particularly notable as the barriers to improve reproducibility through sharing of digital artefacts are rather low. Yet, it must also be noted that not all software and data can always be freely and immediately shared. The claim for reproducibility does not overrule any legal or privacy concerns. Our results apply primarily to datasets generated through experiments without human participants. Future research should investigate incentives and requirements for sharing data from human subject research.

\section{Limitations and Future Work}

We aim to foster the reproducibility of our work and to provide a base for future research. Therefore, this paper is accompanied by various resources from our study. Those include the semi-structured interview questionnaire, the ATLAS.ti code group report and the templates of the two paper exercises. As is the core idea of reproducible research, we envision future work to extend and enrich our findings and design implications by studying perceptions, opportunities and challenges in diverse scientific fields. We can particularly profit from empirical findings in fields that are characterized by distinct scholarly communication and field practices and a differing role of reproducibility. Also different forms of research will need to be studied. Our study's focus is on data-intensive natural science, using the example of computational research in HEP. It does not intend to contribute directly to other forms of research such as descriptive and qualitative research. 

It should also be noted as a limitation of the study that the reference preservation service is based entirely on custom templates. While this does not reflect the majority of repositories and cloud services used today for sharing research, our findings indicate that templates are key to enable and support secondary usage forms. And even though our study focused solely on HEP, findings and implications are however likely to be relevant for numerous fields, in particular computational and data-driven ones. Uncertainty, visibility and automation are of general concern to researchers, with HEP representing an ideal study context that provides one of the most data-intensive, diverse, distributed and technology-adopting environments.

\section{Conclusion}

This paper presented a systematic study of perceptions, opportunities and challenges involved in designing technology that enables research preservation and reproducibility in High Energy Physics, one of the most data-intensive branches of science. The findings from our interview study with 12 experimental physicists highlight the resistance and missing motivation to preserve and share research, core requirements of reproducible science. Given that the effort needed to follow reproducible practices can be spent on novel research --- usually perceived to be more rewarding --- we found that contributions to research preservation technology can be stimulated through secondary benefits. Our data analysis revealed that contributions to a centralized preservation platform can target issues and improve efficiency related to \textsc{communication}, \textsc{uncertainty}, \textsc{collaboration} and \textsc{automation}. Based on these findings, we presented implications for designing technology that supports reproducible research. First, we discussed how studying researchers' practices enables exploiting secondary usage forms of platforms and its content that are expected to stimulate researchers' contributions. Centralized repositories can promote preservation as an effective strategy to cope with uncertainty; support locating expertise in research collaboration; and provide a more guided and efficient research process through preservation templates that closely map research workflows.

\balance{}
\balance{}

\begin{acks}
This work has been sponsored by the Wolfgang Gentner Programme of the German Federal Ministry of Education and Research (grant no. 05E15CHA).
\end{acks}

\bibliographystyle{ACM-Reference-Format}
\bibliography{sigchi}

\end{document}